\documentclass[version=preprint]{iacrcc}
\pdfoutput=1
\newcommand{\aurl}[1]{\url{https://#1}}
\newcommand{\cmd}[1]{\texttt{\textbackslash{}#1}}

\usepackage{xcolor}
\newcommand{\xmltag}[1]{\texttt{<{#1}>}}

\usepackage{todonotes}

\newcommand{\todok}[1]{\todo[color=green!20,inline]{K: #1}}

\newcommand{\BibTeX}{{\rmfamily B\kern-.05em%
    \textsc{i\kern-.025em b}\kern-.08em%
    T\kern-.1667em\lower.7ex\hbox{E}\kern-.125emX}}

\newcommand{\quem}{\tikz[baseline=(wi.base)]{\node[fill=black,rotate=45,inner
      sep=.1ex, text height=1.8ex, text width=1.8ex] {};
    \node[font=\color{white}] (wi) {?};}}

\title[subtitle={(updated April 3, 2023)}]{LaTeX, metadata, and publishing workflows}

\addauthor[orcid   = {0000-0003-1010-8157},
           inst    = {1},
           onclick = {https://www.joppebos.com},
	   email   = {joppe.bos@nxp.com},
           surname = {Bos},
          ]{Joppe W. Bos}
\addauthor[orcid   = {0000-0001-7890-5430},
           inst    = {2},
           email   = {iacrcc@digicrime.com},
           surname = {McCurley},
          ]{Kevin S. McCurley}

\affiliation[ror      = 031v4g827,
             onclick  = {https://www.nxp.com},
             street   = {Interleuvenlaan 80},
             city     = {Leuven},
             postcode = {3001},
             country  = {Belgium}
            ]{NXP Semiconductors}
\affiliation[country={United States}]{Self}
\license{CC-by}

\usepackage{newunicodechar}
\newunicodechar{ế}{\'{ê}}
\begin{document}

\maketitle

\keywords[publishing, LaTeX, metadata]{publishing, \LaTeX, metadata}

\begin{abstract}
  The field of scientific publishing that is served by \LaTeX\ is
  increasingly dependent on the availability of metadata about
  publications. We discuss how to use \LaTeX\ classes and
  \BibTeX\ styles to curate metadata throughout the life cycle of a published
  article. Our focus is on streamlining and automating much of
  publishing workflow. We survey the various options and drawbacks 
  of the existing approaches and outline our approach as applied in a new
  \LaTeX\ style file where we have as main goal to make it easier for authors to specify
  their metadata only once and use this throughout the entire publishing pipeline. 
  We believe this can help
  to reduce the cost of publishing, by reducing the amount of human
  effort required for editing and providing of publication metadata.
\end{abstract}

\section{Introduction}
The original goal of \TeX\ was focused primarily on typesetting, and
the appearance of the output on paper. The later invention of
\LaTeX\ was focused on ``letting the user concentrate on the
structure of the text rather than on formatting
commands''~\cite{latex}. Users were encouraged to write their papers
using high-level macros like \texttt{\textbackslash{}section}, and
leave the decisions like how much space to put before or after a
section to the style that is used. As a result, an author does not
have to worry so much about how the paper looks, but only has to worry
about how the paper is logically structured. This ``separation of
concerns'' is an example of a more general concept from computer
science where a program would be separated into distinct
units that focused on a restricted part of the task for the
program.


The separation of concerns about appearance versus structure has proved to
be very effective and most, if not all, scientific publishers now have
their own \LaTeX\ styles (overleaf lists about 400 different styles for
academic journals.\footnote{See \aurl{www.overleaf.com/latex/templates/tagged/academic-journal}}
These styles make it easy for an author to conform
to a common look and feel in a journal. This can streamline the production
steps for a journal if authors comply with the style.

There is at least one area in
which the \LaTeX\ community has been slow to adapt to the needs of
modern publishing workflows, namely in the {\em curation of metadata
about publications}. This is the main focus of this article.
Some of this metadata is fairly obvious, such as the title, list of
authors, affiliations, funding sources, and references.  Most of this
metadata should be supplied by the authors, who best understand the
relationships between the different entities.  We believe that a
\LaTeX\ class should provide a convenient mechanism for authors to
enter the metadata only once, in a way that encodes relationships
between the entities, so that this \LaTeX\ class can generate appropriate
machine-parsable formats which can be used at every phase in the publishing
pipeline.

In this paper we survey the various requirements for metadata and the 
options one can use to collect this information conveniently for the authors. 
We outline possibilities and drawbacks of some of the \LaTeX\ engines and 
introduce our approach to solve this problem in a new \LaTeX\ class file.  

\section{The importance of metadata}\label{sec:importance}
In the increasingly competitive world of academic publishing, there
are several reasons why metadata has become important:
\begin{itemize}
\item Metadata provides an important taxonomy for navigating in the
  world of academic literature.
\item Authors, institutions and funding agencies need to receive proper
  attribution.
\item The proliferation of publications has increased the need to rank
  publications, authors, universities, and departments.  The
  primary methodology for measuring scientific impact and ranking of
  publications is based on citation analysis, to determine which
  publications have the greatest influence on (an area of) science.
  Researchers themselves need to focus their attention on the most
  important works in a field.  Quality metrics can help, even if they
  are imperfect.
\item Papers often end up being posted in multiple places on the web
  in various forms, so duplicate elimination is a serious
  problem. Publishers often exacerbate this problem by trying to keep
  an ``officially published'' version behind a paywall, while allowing
  authors to freely post unofficial versions on their home pages and
  eprint servers.
\end{itemize}
Academic publications are now being tracked by numerous indexing
services including Google Scholar, Clarivate (Web of Science), Scopus,
PubMed, and many others.  Their task is complicated by the poor
state of metadata within publications. The problem of inferring
metadata from publications is itself a complicated
subject~\cite{arxiv.2006.05563} that could have been simplified if the
metata had been provided with the publication itself in a
machine-parsable format. 
This is one of the goals we set out to achieve in this work. 

\section{Publishing workflow}
Part of our motivation arises from our involvement in trying to launch
a new open access journal for the professional non-profit society
International Association for Cryptologic Research 
(IACR).\footnote{See \aurl{iacr.org}.}
The society already runs two
diamond open access journals, but experience from running these has
shown that on average each published paper requiries about an hour of
human effort for production and metadata handling.

Another
study~\cite{f1000} estimated the amount of human labor for editing and
production to be 7.5 person-hours for each published paper. We believe
that most of this should and could be automated, and this can help to lower the
cost of publishing.  This is particularly important for open access
publishing, which is heavily dependent on volunteer
labor~\cite{bosman_jeroen_2021_4558704} as a way to control costs. It
can also be used to improve the profitability of commercial publishers.

Publishing workflows vary somewhat from one organization to another,
but will at least include the steps of authoring, submission, review,
revision, submission of a final version, copy editing, production of
readable output formats (e.g., PDF and HTML), DOI registration,
indexing, and hosting. Metadata can be important even during the
reviewing phase, because it can be used to automatically identify
potential conflicts of interest. For large scale operations such as
the 2022 NeurIPS conference that received 9,634 full paper submissions with
10,406 reviewers,\footnote{See \aurl{media.neurips.cc/Conferences/NeurIPS2022/NeurIPS_2022_Fact_Sheet.pdf}.}
the metadata can help greatly in automating assignment of reviewers and avoiding 
conflict of interests.

In some systems like Open Journal Systems~\cite{ojs}, the submission and curation of
metadata is treated as a separate task from submission of the Word,
\LaTeX\ or PDF document. That imposes an extra burden on authors, but
also renders the workflow vulnerable to inconsistencies with metadata
in two places. In our experience, by the time an article has been
revised and accepted, there are often changes in titles, abstracts,
affiliations, email addresses, etc. Checking and correcting these inconsistencies
ends up costing human time by authors and editors.

Metadata becomes increasingly important in the publishing workflow when the
author submits their final version. Once the final version is ready,
it will typically be registered for a DOI, and this requires supplying
a considerable amount of metadata. Moreover, the web ``landing page''
for a paper typically has to be created from the metadata. Indexing agencies
then step in, either by crawling the data or by metadata feeds from the
publisher.

\section{Our approach at a high level}
We automate the capture of metadata during the publishing workflow
through the use of a \LaTeX\ class \texttt{iacrcc.cls}
and a \BibTeX\ style \texttt{iacrcc.bst}.%
\footnote{The authoritative place to download these is \aurl{publish.iacr.org/iacrcc}.}
\begin{itemize}
\item Authors prepare their \LaTeX\ in the usual fashion using the
  provided \texttt{cls} file. Almost all metadata is supplied by authors using
  structured macros from the class file. The function of the \texttt{cls} and
  \texttt{bst} files
  is to both display the metadata in the output format, but to also
  extract the metadata during the compilation process, producing an easily
  parsable external format as a side product.
\item When authors supply their final versions, they do so by uploading
  their \LaTeX\ source to a cloud server, which compiles their sources
  and extracts all metadata from their sources into a structured format
  text file. The submission process does
  not require authors to enter any additional metadata, because it is
  all encoded into the \LaTeX\ source. The DOI suffix is assigned by the server and
  the DOI is compiled directly into the PDF at time of submission (the DOI will
  be registered with the DOI registration agency later).
\item A post-compilation step is used to parse the structured metadata
  and convert it into other formats, including JSON and XML.
\item Once the paper has been approved by an editor, the extracted
  metadata is further used to register the DOI. The extracted metadata
  is also used to produce various web pages for the journal site, RSS
  feeds, OAI-PMH feeds, and register with various indexing services.
\end{itemize}
The metadata output we require is necessarily {\em text}, and the
lingua franca of text encoding is UTF-8. With the exception of
mathematical structures like inline equations in titles or abstracts,
this text should be devoid of \TeX\ macros. This causes a few problems in
the \LaTeX\ world.

In our discussion, it is important to distinguish between text
\emph{input} encodings and text \emph{output} encodings in the
\TeX\ world. Traditionally, \TeX\ was written to accept only 7-bit
ASCII input, using things like |verb|\"u| to encode accented
characters. All \TeX\ engines are now capable of dealing with UTF-8
\emph{input} encodings, but the support for \emph{output} encodings is
much more complicated. The problem here is fonts, which are of course
necessary for typesetting. \TeX\ engines are generally concerned with
producing typeset material with fonts and hyphenation, and this
overloads all discussion about \TeX\ output encoding. \LaTeX\ also has
an internal representation for characters called the ``LICR''.  A
complete discussion of this is beyond the scope of the paper, but
see~\cite{encguide} and~\cite[§ 7]{latexcompanion} for more details.

Some characters like \texttt{\# \$ \% \& \_ \^{}
  \textasciitilde\ \textbackslash\ \{ \} \%} and spaces
require special effort by
authors to encode these as text, because they have special meaning to
the \TeX\ parser. When \LaTeX\ reads an input file, the contents are
converted to a sequence of tokens, some of which represent text
characters. This internal representation makes it inconvenient to
access what would commonly be thought of as ``text'' in any other
computer language.  Moreover, the core functionality of \LaTeX\ is
user-defined macros, so an author might define \cmd{pe} to
represent the text string ``Paul Erdős''.  We only discover this
during the \LaTeX\ output process when the macros are expanded into
tokens. The \TeX\ expansion process is very
complex and has many special cases.

\section{Metadata capture}
We considered several ways to implement metadata extraction from
author's \LaTeX\ sources.
One approach would be to use a \LaTeX\ parser in a high level
programming language to extract the metadata
directly from the \LaTeX. The problem of parsing \LaTeX\ is
complicated by the need to expand macros, for which the
\LaTeX\ engines themselves are so far the only robust
solution. Another approach that we considered involved using
\texttt{lua} within \texttt{lualatex}. This is much better suited to
text processing than using \LaTeX\ itself, but we had an initial goal
to try and make things work with any \LaTeX\ engine. The third
option that we tried uses the \cmd{write} macro to capture metadata
and write it out to a file.

\subsection{Our first attempt with \cmd{write}}
The function of the \cmd{write} macro is to expand a list of tokens
and write a parsable representation of these tokens into a file.  It
was originally intended for the writing of auxiliary files like
\texttt{.bbl}, \texttt{.aux}, etc that would be read into the
\LaTeX\ source during subsequent compilation. The fact that
\cmd{write} performs expansion is very useful to us, because it
expands user-defined macros. Unfortunately we found that \cmd{write}
has a few problems when we use it to produce metadata.

\begin{enumerate}
\item The \textasciitilde\ character gets expanded by \cmd{write}
  into the following unfriendly token representation:
\begin{verbatim}
  \protect \unhbox \voidb@x \protect \penalty \@M \ {}
\end{verbatim}
  From the point of view of metadata, it should be
  replaced by a non-breaking space, which is present in essentially
  every character encoding.
\item The expansion process sometimes produces extra spaces, e.g.,
  \verb+$\alpha$+ becomes \verb+$\alpha $+. More seriously, if we
  change the \verb+$+ math delimiters to the \LaTeX\
  versions \verb+\(+ and \verb+\)+, then \cmd{write} fails completely,
  and part of the errors go to the output file. There are other
  examples of things that cannot be provided as arguments
  to \cmd{write}, including macros with optional arguments.
\item The expression \verb|$\frac{1+x}{2-x}$| gets expanded into the less
readable version
\begin{verbatim}
  $\protect {\begingroup 1+x\endgroup \over 2-x}$
  \end{verbatim}
\item Text processing in \LaTeX\ is conflated with typesetting,
and it's possible to get mixed encodings in the output. This is acceptable
as an intermediate step in typesetting, because \TeX\ translates this
to glyphs on the page using fonts encoded in different encodings.
Unfortunately it is unacceptable for processing of metadata from \cmd{write}.
If we use \verb+\write\f{Ð and \DJ\ and ü and \"u}+ then
the output from \texttt{lualatex} consists of:
\begin{verbatim}
Ð and \TU\DJ \ and ü and ü
\end{verbatim}
because it uses a unicode encoding, but the output from pdflatex consists of:
\begin{flushleft}
\texttt{Ð and \textbackslash{T1}\textbackslash{DJ} \textbackslash\ and ü and }\quem
\end{flushleft}
(the \cmd{DJ} macro requires us to use \verb+\usepackage[T1]{fontenc}+). Unfortunately,
this output from \cmd{write} contains characters with mixed encodings: The Ð is encoded as
UTF-8 \verb+C390+ and the ü is also correctly encoded, but the \verb+\"u+ is encoded as
the T1(Cork) encoding of \verb+FC+ rather than the UTF-8 encoding of \verb+C3BC+. 
\end{enumerate}

Many of these things could be fixed in a post-processing step of our
pipeline written in a higher level language, but we are unable to
tolerate having multiple character encodings in the output
file. Moreover, these are merely examples of some of the things we
might encounter in an author-supplied \LaTeX\ that compiles correctly.
Our goal is to provide a system that supports whatever the author
supplies to us, and to provide them with clear instructions on how to
prepare it without causing any interruptions (errors) or other
inconvenience to the author's typesetting experience. Ultimately the
metadata that the author encodes as text, either as
\TeX\ macros or as UTF-8 input should survive as UTF-8 output.

One suggestion we received was to try using the \texttt{expl3} macro
\cmd{text\_purify:n}, and we found that this is useful to clean up a few
things like removing \cmd{texttt}. Unfortunately we also found that it
performs some destructive things like completely removing some macros like
\cmd{dag}, \cmd{copyright}, and \cmd{pounds} that produce
\dag, \copyright, and \pounds\ with obvious encodings in UTF-8 as U+2020, U+00A9,
and U+00A3.  We were also unable to understand the behavior of this
function from the documentation or the source code, so we decided to
stop using it. Perhaps others will find it useful.

There are other \LaTeX\ packages that have required encoding of text. A good
example is the \texttt{hyperref} package, which needs to produce PDF
bookmarks encoded in a text encoding supported by PDF. PDF originally
supported only the PDFDocEncoding text encoding. Starting in 1996, PDF
supported the UTF-16BE encoding, which is the UTF-16 encoding with a
byte order marker at the beginning of the string. The UTF-16 encoding
is quite different than UTF-8, since it consists of one or two 16-bit
blocks per character, whereas UTF-8 consists of between 1 and 4 bytes
to represent characters.  Starting with PDF~2.0 in 2017, text may be
encoded in PDF as UTF-8, but as of 2022, the \texttt{hyperref} package
still supports only UTF-16BE (PU) and PDFDocEncoding formats for
text. The unicode implementation depends on a 2000-line file with
lines like
\begin{Verbatim}
\DeclareTextCompositeCommand{\`}{PU}{u}{\80\371}% U+00F9
\end{Verbatim}
to define the encoding a 'Latin Small Letter U with Grave' ù.
Building such a thing is a tedious process.

Another example of a \LaTeX\ package with requirements for text
processing is the \texttt{pdfx} package that produces UTF-8
encoded XML inside the PDF output from \LaTeX. They defined a
pseudo-encoding called \texttt{l8u}~\cite[§ 4.2]{pdfx}. They mention
that such an encoding could be repackaged into a full package to
produce UTF-8 encoded strings, and that might have been useful to us.
The \texttt{pdfx} package's purpose is similar to ours, in that their
objective is to \emph{inject} a subset of metadata into the PDF. Our
goal is to \emph{extract} a larger set of metadata from the
\LaTeX\ sources (e.g., ORCID IDs).

\subsection{Switching to \cmd{protected@write}}
After an earlier version of this paper, we received comments from
reviewers that suggested
using the \LaTeX\ macro \cmd{protected@write} instead of the \TeX\
macro \cmd{write}. This solves a number of problems for us. For
example \verb+\(+ is allowed within \cmd{protected@write}, and it solved
the problem of mixed character encodings with \texttt{pdflatex}.

Unfortunately the use of \cmd{protected@write} introduces some other
oddities such as expanding \verb|\frac{x}{2}| to
\cmd{protect}\textvisiblespace\cmd{frac}\textvisiblespace\textvisiblespace\verb|{x}{2}|.
Luckily we don't really need the output file to contain only UTF-8
characters, because we can easily use a postprocessing step in a
higher-level language like python to remove the spurious \cmd{protect}
macros, or convert something like \verb+\"u+ to \"u, and this is what
we now use. It remains to be seen what problems arise from other author-supplied
\LaTeX.

\section{What metadata is required?}
There is potentially a lot of metadata that can be associated with a
publication. Some fields are quite obvious, but even the obvious fields have
nuance in how they are encoded. Examples include:
\begin{itemize}
\item Title of the work. In some fields it is commonplace to
  use mathematics in titles, but \TeX\ formatting in metadata records
  is often changed to another format like MATHML. Titles may also
  encode face markup (e.g., bold face) or multiple character sets.
  Extremely long titles are sometimes broken up into a hierarchy,
  incorporating a subtitle or short versions for running titles.
  An article may have titles in different languages.
\item Authors of the work. Just asking for names of authors can be tricky, and
  readers are urged to read \cite{falsehoods}. About the best one can hope for is
  to provide a string field for a name, but some bibliographic formats may insist upon
  entering a surname and/or given name.
\item
  Authors may have different levels of contribution. In some cases this is signaled
  by having author names out of alphabetical order, but in other fields it is
  common to identify a \emph{role} for author contributions. The {CRediT}
  taxonomy is often used to reflect this~\cite{credit}. Authors may also be categorized as
  a  ``corresponding author''.
\item Relationships between authors and affiliations and/or
  authors and funding agencies. It is now very common for authors to
  have multiple affiliations~\cite{hottenrott2021rise} and multiple authors 
  to share a subset of affiliations or funding agencies. These many-to-many
  relationships are best encoded as relations rather than repeating the
  information for each author. 
\item Bibliographic information (e.g., journal or conference name, volume,
  year, etc).
\item A list of references (supplied from {\BibTeX}).
\item Licensing information.
\end{itemize}
  
There are numerous other fields that may be encoded into a
\LaTeX\ document or the output format produced from \LaTeX{}.
Examples include abstract, number of pages, address information for
authors, email address of authors, links to ancillary works like code
and data, etc. It is beyond the scope of this document to catalog all
of them, but there have been numerous attempts to define schemas for
publishing metadata. We mention them here because the reader may have
a requirement to comply with one or more.

Metadata can of course be encoded into various formats such as the
\LaTeX\ itself, XML, JSON, or YAML. Most of the descriptions of
publication metadata is done in various flavors of XML. We believe
that author-supplied metadata should be encoded into the original \LaTeX{},
and that processing of this metadata should be capable of producing
other formats that are required for the publication pipeline.

Below we describe a few of the most important metadata schemas.

\subsection{Crossref}
\href{https://crossref.org}{crossref.org} 
  is a non-profit organization whose primary mission is the
  collection of metadata and the assignment of DOIs.  As such, they are
  one of the most important formats to be compatible with, and they have
  own XML schema~\cite{crossrefschema}.
  Their schema supports multiple affiliations, author roles, and funding
  agencies.

  \subsection{JATS}\label{subsec:JATS}
  The Journal Article Tag Suite (JATS) is an XML format that has three
  variations for archiving \& metadata, publishing, and
  authoring~\cite{jatsguide, JATSschema}. The JATS format may be viewed as a
  complete structural representation for a publication; in many ways
  comparable to \LaTeX\ but focused more on structure and less on
  layout.  A JATS document consists of several sections, including:
  \begin{description}
  \item[front matter] This is a required section where most important metadata
    about the article will appear. It contains metadata about the article
    but also about the containing serial or journal.
  \item[body] This optional section is indended to contain the main content of the article itself,
    and consists of sections, paragraphs, etc.
  \item[back matter] This optional section contains bibliographic references,
    a glossary, appendices, or other ancillary information for the article.
    One area in which JATS excels is in the encapsulation of bibliographic references.
  \end{description}
  The schema for front matter is very expressive and offers the
  broadest coverage of publishing metadata that we have found. JATS can accomodate both
  MATHML3 and inline \TeX\ for mathematics, and as such may also be
  considered as a competitor to \LaTeX{}, or as an output format defined
  in section~\ref{outputformats}.  It is however quite difficult to
  convert the full capabilities of \LaTeX\ into JATS, and the results
  are often not faithful to the original intent of the author because
  JATS contains only semantic structure rather than layout information. In order
  to convert JATS to a convenient consumable format, it requires a counterpart
  to \LaTeX\ styles, namely XSLT or some other means of formatting the XML.
  
  \subsection{Elsevier's Scopus}
  Scopus is one of the largest indexing agencies. As a commercial
  service, they  partner with publishers to receive their content, but their
  preferred method is to receive PDF and XML in JATS format~\cite{scopusfeed}.

  \subsection{Clarivate Web of Science} Clarivate's Web of Science is another commercial
  service that provides indexing services. They accept XML feeds in their own
  schema~\cite{clarivate}.

\subsection{Other formats}\label{subsec:Dublin}
There are a number of less descriptive metadata formats, but they are not as
comprehensive. We mention them in part because they may define namespaces
that can be used within other XML formats.
\begin{description}
\item[{DOAJ}] DOAJ stands for the Directory of Open Access Journals,
    which is is a community-curated online directory that indexes open
    access peer-reviewed journals. They accept XML in their own schema
    or crossref's schema~\cite{doaj}.
  \item[Extensible Metadata Platform ({XMP})] This is an ISO standard
    for representation of metadata associated with digital
    documents. It was originally created by Adobe Systems and is often
    used in {PDF}.  It defines several ``standard'' schemas, but is
    extensible through inclusion of other {XML} schemas.
    The standard schemas lack many basic features that are important
    to describe academic literature.
  \item[Dublin Core] The Dublin Core Data set version 1.1 started in
    1999 as an XML schema consisting of only fifteen basic
    elements. It has proved to be inadequate to describe scholarly
    articles, but some of the elements are still used in other XML schema, and
    it's one of the standard schemas for XMP.
    Version 1.1 was improved upon in the ``Qualified Dublin Core'' in 2008,
    but this still lacks basic things like ORCID IDs or affiliations. Bibliographic
    references are encoded as an unstructured string 
    and omitted entirely in Dublin Core 1.1.
  \item[{PRISM}]
    PRISM~\cite{prism} is another simple XML format to describe authors,
    title, url, DOI, dates, rights, etc. It seems to be less widely used than
    {JATS}. The \texttt{hyperxmp} package generates XMP in PDF with
    some tags from this schema.
  \item[{TEI}] The Text Encoding Initiative ({TEI}) is an older
    XML-based format that started in 1987. It is used mainly in the
    humanities as an alternative to {JATS}, but is much less
    descriptive.
\end{description}

\subsection{Conclusions about metadata schemata}
When developing a \LaTeX\ class for a journal and/or conference
series, it's important to be aware of the metadata elements that are
most relevant to the field. Since our journal was in computer science, engineering and
mathematics, we had no need for things like clinical trial information
or chemical reaction encoding.  A complete \LaTeX\ package that
supports all possible metadata for a publication would be a daunting
project, and would impose an undue burden on authors to wade through
all the possible metadata elements.

Our workflow supports registration of DOIs with crossref and indexing
by several services.  Compliance with these schemas imposes some
restrictions on the relationships between articles, authors,
affiliations, and funding agencies. These are shown in
Figure~\ref{fig:relationships}. We believe these are fairly universal
to all disciplines, but there are still many potential extensions. The
entities like author or affiliation can be extended through attributes
on the entities themselves, or on the relationships between them. As
an example, an author may list an affiliation but say that they were
on leave when the work was done. This would be an attribute on the
relationship between an author and the affiliation. An author may be
listed as deceased, with an attribute on the author. Many such
attributes are described in the {JATS} schema, and some are linked via
footnotes with a footnote type. We defer such things to future work.

\usetikzlibrary{shapes.geometric,arrows.meta,bending}
\begin{figure}
  \begin{center}
    \tikzstyle{entity} = [rectangle, rounded corners, minimum width=1.5cm, minimum height=.6cm,text centered, draw=black, fill=green!30]
    \tikzstyle{author} = [rectangle, rounded corners, minimum width=1.5cm, minimum height=.6cm,text centered, draw=black, fill=blue!30]
    \tikzstyle{affil} = [rectangle, rounded corners, minimum width=1.5cm, minimum height=.6cm,text centered, draw=black, fill=red!30]
    \tikzstyle{funder} = [rectangle, rounded corners, minimum width=1.5cm, minimum height=.6cm,text centered, draw=black, fill=yellow!30]
    \tikzstyle{arrow} = [thick,->,>=stealth]
    \begin{tikzpicture}[node distance=2cm]
      \node (article) [entity] {Article};
      \node (author1) [author,below of=article] {Author 1};
      \node (author2) [author,below of=article,yshift=-1cm] {Author 2};
      \node (funding1) [funder,right of=article,xshift=2cm] {Funding 1};
      \node (funding2) [funder,right of=article,xshift=2cm,yshift=-.8cm] {Funding 2};
      \node (affiliation1) [affil,right of=author1,xshift=2cm] {Affiliation 1};
      \node (affiliation2) [affil,right of=author1,xshift=2cm,yshift=-.8cm] {Affiliation 2};
      \node (affiliation3) [affil,right of=author1,xshift=2cm,yshift=-1.6cm] {Affiliation 3};
      \draw [arrow] (article) to (author1);
      \draw [arrow] (article.south) to [out=210,in=-210] (author2.west);
      \draw [arrow] (article) -- (funding1);
      \draw [arrow] (article) -- (funding2.west);
      \draw [arrow] (author1) -- (affiliation1);
      \draw [arrow] (author1) to [out=-10,in=-210] (affiliation3.west);
      \draw [arrow] (author2) -- (affiliation1);
      \draw [arrow] (author2) -- (affiliation2.west);
      \draw [arrow,dashed] (author1) -- (funding1.west);
      \draw [arrow,dashed] (author2) -- (funding2);
    \end{tikzpicture}
    \caption{Relationships between major entities. Each entity is listed
      only once in the article.  An article may have multiple authors
      who share relationships to affiliations.  Funding agencies are
      related to the article in the crossref schema, so we chose to
      link them this way. As an alternative, relations shown with dashed
      arrows can link authors to their funding sources, in much the same
      way that we relate authors to their affiliations. We chose to use
      footnotes to clarify the complex relationships between funding agencies
      and authors or  affiliations. Some funding agencies (e.g.,~\cite{nihfunding})
      have strict guidelines
      for how these annotations should be shown in the paper.
      \label{fig:relationships}}
  \end{center}
\end{figure}

\section{Using unique identifiers}
Unfortunately, things like titles and human names are not unique
identifiers.  DBLP lists 14 authors in computer science who use the
exact name ``Thomas M\"uller'', and dozens of others that are similar
to this, like Thomas F. M\"uller.  In order to perform large scale
bibliometric analysis for attribution or duplicate detection, all
entities associated with a publication need to be assigned a unique
identifier.

Many of the {XML} schemas like {JATS} have embraced the use of
unique identifiers. The most notable efforts to assign unique identifiers
include:
\begin{itemize}
\item DOIs for publications~\cite{paskin2010digital},
\item ORCID IDs for authors~\cite{haak2012orcid},
\item ROR IDs for research institutions~\cite{ROR},
\item Crossref funder registry for funding agencies~\cite{funder}.
\end{itemize}

Note that in each case where an organization has assigned a unique ID
to an entity, there will often be competing organizations with their
own ID space. For example, DOIs cost money, so some organizations have
assigned IDs for articles within their own namespace. For example,
\texttt{arxiv.org} has long been assigning unique IDs to the papers in
their namespace. They recently also started assigning DOIs that encode
their long-standing ID system.

Sometimes two nearly identical versions of an article will have two
DOIs assigned to them in the same namespace (e.g., the publisher
version and the arXiv version). This is not the only case when
multiple DOIs might be assigned to minor variations of a
paper. \texttt{arxiv.org} assigns a different DOI to each revision, no
matter how small.

ORCID IDs are actually a reserved block of ISNI identifiers, but ISNI
has also issued IDs to some 
authors\footnote{See for example \aurl{isni.org/isni/0000000088556337}.}
who don't have ORCID IDs. ISNI also maintains identifiers in
other domains, including creators of media, publishers, and
some academic institutions\footnote{See for example \aurl{isni.org/isni/0000000121924307}.} 
as an alternative to ROR IDs. Commercial services that
track bibliographic information will typically maintain their own
namespace for author and/or affiliation identifiers.  For example,
other identifiers for authors have been issued by Clarivate Web of
Science, Scopus, SciENcv, Mathematical Reviews, and DBLP.

ROR IDs have coarse granularity, so while there is an identifier
for Massachusetts Institute of Technology, they don't distinguish
between departments, schools, or programs of the university. By
contrast, Mathematical Reviews assigns institution codes at the
department level (e.g., {1-SCA-C} for the department of computer
science at University of Southern California).

A complete list of identifiers associated with scholarly publications
is beyond the scope of this document, and we should probably expect
future ID systems to emerge.  Because an entity may have multiple IDs
from different organizations, we strongly recommend a schema that
assigns IDs with a namespace and identifier within that
namespace. Thus for example, a paper on \texttt{arxiv.org} might have
an identifier within the DOI namespace, but also an identifier within
the arxiv.org namespace. The ability to include identifiers from
multiple namespaces can help with disambiguation.  

\subsection{Major takeaways about identifiers}
There are a {\em lot} of potential metadata elements that may be
associated with a scholarly article, and the list continues to grow.
So long as data is collected in the most granular way possible, it
should be possible to reformat it into whatever XML schema is desired.
We believe that the JATS format (see Section~\ref{subsec:JATS}) offers
the most descriptive language for metadata about publications, and we
have designed our metadata macros to align with JATS.

Different participants in the publishing world will have different
required elements and optional elements. It is important to collect as
much information as possible in order to maximize compliance with
downstream consumers of metadata.  \LaTeX\ styles tend to be written
for specific journals, societies, and conferences, and they need only
embrace the metadata elements that are most common to their discipline
(e.g., clinical trial data is irrelevant in mathematics). Finally, you
should probably expect that the schema may need to be extended in the
future.

\section{Output formats}\label{outputformats}
An important step in the publishing workflow is to create various formats
for human consumption (e.g., PDF and HTML). To the extent possible, it is desirable
to embed the metadata into these output formats in a machine-readable way
so that the metadata accompanies the consumable document.
Unfortunately the standards for doing so are generally lacking in comprehensiveness.

\subsection{Embedding metadata in {HTML}}
{HTML} has very incomplete specifications for embedding metadata.  The
original \xmltag{meta} tag was intended for metadata, but this is a
flat structure, and the original \texttt{name} attribute covers
relatively few of the metadata elements for a scholarly publication.
There are various extensions that have been proposed for HTML,
including the
whatwg list\footnote{See \aurl{wiki.whatwg.org/wiki/MetaExtensions}.} 
of names for \xmltag{meta} tags. Unfortunately this still lacks many of
the elements we need.

An alternative would be to
encode XML within \xmltag{script type="application/xml"}. This would work if
your XML does not contain \xmltag{/script}. Another option is to use
\begin{quote}\xmltag{link itemprop="meta" src="xmlurl"}\end{quote} or
\begin{quote}\xmltag{link rel="alternate" src="xmlurl"}\end{quote}
in the \xmltag{head} of the document. Neither of these are standard.

Another schema for bibliographic information in a web page
was provided by the 
ScholarlyArticle format\footnote{See \aurl{schema.org/ScholarlyArticle}.}
from \texttt{schema.org}. This is inadequate for a
number of reasons, including the fact that it doesn't contain ORCID
IDs. Theoretically it is possible to extend ScholarlyArticle
in a way that patches the schema deficiencies, but they have failed to do so
and this appears to have very little usage.

Metadata for \texttt{schema.org} is typically embedded in a page as
\texttt{JSON-LD} in the \xmltag{head} of the HTML page as
\begin{quote}\xmltag{script type="application/ld+json"}
\end{quote}
One approach that we find
promising would be to embed JATS in this way, by converting the XML to
JSON. Unfortunately there is
\href{https://www.p6r.com/articles/2010/04/05/xml-to-json-and-back/}{no standard way} 
of converting XML to JSON.  There is clearly a need for
standardization in order for this to be useful.

\subsection{Embedding in PDF}
While we should expect other consumable document formats to grow in
popularity in the future, today's usage of \LaTeX\ is primarily to
produce PDF. There are at least two ways to embed metadata into PDF,
namely the Document Information Dictionary and XMP. The Document
Information Dictionary has existed since version 1.0 of PDF, but
encodes only very basic elements like Title,
Author, Subject, Keywords, CreationDate, etc. This is the output from
the \texttt{hyperref} package, but this schema is inadequate for the
needs of publishing. In particular it has a single field for author,
and does not have any facility to embed unique identifiers like DOI or
ORCID IDs. Due to these inadequacies, most metadata should be encoded
using XMP, which is described in the next section.

\subsubsection{XMP}
Starting with PDF version 1.4, the preferred way to embed metadata
into PDF is as ``Metadata Streams'', which are themselves XML
documents called packets.  This is referred to as the Extensible Metadata Platform
(XMP)~\cite{xmp}.  The XMP standard defines several standard
schemas~\cite{pdfaxmp}, but the only one that appears to be 
relevant for academic publishing is minimal Dublin Core Schema 1.1
that we already dismissed as inadequate (see Section~\ref{subsec:Dublin}).  
Luckily, as the name
implies, this format is extensible, and the XML dictionary may use a
schema from a variety of namespaces~\cite{xmpextension}. The
\texttt{hyperxmp} package already does this for a few metadata elements.

We believe that there is considerable room for improvement in the use
of XMP metadata, including the encoding of bibliographic references,
ORCID IDs, affiliations, funding information, etc. Unfortunately the
XMP extension mechanism requires that if any extension schema is used,
then the entire extension schema must be embedded into the PDF
document. This can result in a PDF file that has more schema
information than actual metadata.



\subsubsection{PDF/A and archiving}

PDF/A is a set of ISO standards~\cite{pdfa} that define a subset of
PDF, intended for long-term archiving of electronic documents. The
standards define different versions and different levels of
compliance. PDF/A-1 is based on PDF version 1.4; PDF/A-4 is based on
PDF version 2.0.  PDF/A-3 added a feature for file attachments. XMP
for metadata was introduced in PDF/A-1 and is a requirement for
PDF/A-2. Unfortunately, generating a PDF file that adheres to one of
the PDF/A standards can be difficult for a number of reasons:
\begin{enumerate}
\item there are very few open source tools to validate PDF/A
  compliance, perhaps due in part to the fact that the standards are
  proprietary ISO documents that are not freely available. Probably the best open
  source tool is veraPDF, but this is still in beta.
\item Fonts need to be embedded into the PDF. \LaTeX\ engines will usually
  do this automatically, but authors should beware that if they include PDF graphics,
  then they may contain fonts that are not embedded.
\item Color spaces must be included in the PDF. This can be
  problematic if the PDF includes color graphics in different color spaces, because
  only one color space is allowed in PDF/A-1.
\item Certain PDF features such as transparency,
  encryption, embedded audio, and javascript are forbidden.
\end{enumerate}
One requirement for XMP metadata is that the information contained there must
agree with the data in the Document Information Dictionary. This has been pointed out
to be problematic, since the Document Information Dictionary is vague on how to encode
multiple author names~\cite[p. 13]{hyperxmp}.

Some funding agencies such as the National Science Foundation, the
European Commission, and members of {cOAlition S} now require research
publications that have been funded by them to be deposited in approved
repositories for long-term digital archiving.  In particular, NSF now
requires deposit of PDF/A files for all of their sponsored research~\cite{nsf}.
Different repositories will have different requirements for deposit, but PDF/A and JATS are
commonly supported.

It should be noted that the archival value of PDF/A has been
questioned by some~\cite{klindt2017pdf}.

\subsubsection{\LaTeX\ packages for XMP and PDF/A}
Note that the \texttt{hyperref} package includes an option to
generate PDF/A, but this only disables some things that would violate
the PDF/A specification, and does not assist with XMP metadata.
There are several \LaTeX\ packages designed to inject XMP data into PDF
files and/or produce PDF/A compliant output, including
\texttt{hyperxmp}, \texttt{xmpincl}, and \texttt{pdfx}.  These differ
in their goals, their approach and the results. \texttt{hyperxmp} is
incompatible with \texttt{pdfx}. Note also that \texttt{pdfx} requires
the \texttt{luatex85} package to work with \texttt{lualatex}.

The simplest package for authors is \texttt{hyperxmp}. This
modifies the behavior of the \texttt{hyperref} package to generate XMP
metadata in the PDF.  This is used by \texttt{acmart} to provide
XMP, and the \texttt{hyperxmp} package works with \texttt{acmart} to intelligently
infer some metadata.  Unfortunately, \texttt{hyperxmp} has a few problems:
\begin{enumerate}
\item We found that some output fails validation as PDF/A under the
  \texttt{verapdf} package, even under the relatively lax PDF/A 1-b
  standard.
\item It is restricted to a predetermined set of schema, including the
  basic schemas but adding PRISM and others. It omits things like
  ORCID IDs, affiliations, ROR IDs, bibliography, and other elements
  that we consider desirable.
\end{enumerate}

The \texttt{pdfx} \LaTeX\ package~\cite{pdfx} was designed to provide
support for generation of PDF/A compliant documents, and therefore
also supports injecting XMP metadata into a PDF.  It supports
generation of embedded XMP through mappings from \LaTeX\ macros to XML
elements, such as \cmd{Title} to generate \xmltag{dc:title} and
\cmd{Volume} to generate \xmltag{prism:volume}.  Unlike
\texttt{hyperxmp}, the \texttt{pdfx} package requires the author to
specify metadata elements in an external file
\texttt{{\textbackslash}jobname.xmpdata}.  By using these macros, it
can ensure that the data in the Document Information Dictionary and
the XMP data agree.

While \texttt{hyperxmp} and \texttt{pdfx} both support XMP by defining
macros to target specific metadata elements, a more flexible approach
is provided by the \texttt{xmpincl} package.  \texttt{xmpincl} assumes
that you will create the raw XML/RDF file, and just includes
that. This places a considerable burden on the user, but provides
more flexibility than \texttt{hyperxmp}.  The \texttt{pdfx} package
uses \texttt{xmpincl} to include XMP. There is no longer an active
maintainer for \texttt{xmpincl}, but the package itself is fairly
simple.

We should mention that the \LaTeX\ project is working on a
multi-year project for tagged PDF output that defines a new metadata
interface~\cite{latexproject}. That will allow authors to specify a
\texttt{pdfstandard} flag to include a color profile for PDF/A
compliance.  This is being included in \LaTeX~\cite{l3pdfmeta}, but
it still doesn't solve the problem of making sure that included
graphics will comply with the declared color standard. The \LaTeX\ team
is engaged in a long term ``Tagged {PDF} Project'' that has
better support for {XMP} as one goal~\cite{xmp2023}.

\section{\texorpdfstring{\cmd{author}}{\textbackslash{}author} considered harmful}
We now turn to the problem of how to embed metadata into the original \LaTeX\ source.
The original \LaTeX\ definition of \cmd{author} provides little help
in capturing author metadata, and is also problematic for displaying
large numbers of authors. In the standard \texttt{article} class, the
author defines \cmd{author} to include blocks of formatted text,
separated by \cmd{and}.  Thus for example, there is no standard way to
associate an ORCID ID with an author's name, or to associate
affiliations or funding agencies with an author. Left to their own
devices, authors might use various embedded macros or footnotes to
link authors to their metadata, and this makes it very difficult to
extract metadata from the \LaTeX.

Part of the problem here is that the \cmd{author} macro is intricately
woven into the {\em display} of author information on the page. This
is an example where the separation of concerns has been neglected,
mixing structure with display.  Because of this past history with the
\cmd{author} macro, we deliberately chose to break \cmd{author} and
use \cmd{addauthor} instead. This means authors have to do some work
to convert from other standard \LaTeX\ classes to our class, but we
judged that to be necessary because of the bad habits that \LaTeX\ has
encouraged.

The display of author information on an article can be very complex,
particularly when each author may have multiple affiliations, funding
agencies, and footnotes, or when there are a huge number of named
authors.\footnote{e.g., \aurl{arxiv.org/abs/2210.03375}}
The decision of how to {\em display} author information should be a
separate decision from how to {\em capture} the author
information. The role of the author is to correctly specify the
metadata about authors, and the responsibility of the style is to
decide how to lay it out on the page according to the style of the
journal. Note that a journal may also impose special requirements
on how the author metadata is encoded (e.g., whether to accept
pseudonyms or ghost authorships, whether authors are grouped by roles,
what kind of contact information is required, etc).

We are not the first to have recognized this problem, and some
\LaTeX\ styles have improved upon the basic use of \cmd{author}, and
have adopted metadata capture as part of their authoring process.

\begin{description}
\item[authblk] This package provides several ways of displaying author
  information; either as footnotes or as blocks of text.
\item[ltugboat] The style used by TUGBoat redefines the \cmd{author}
  macro to capture an individual author's name. Sequential calls to other
  macros like \cmd{ORCID} allow linking of metadata to individual authors.
\item[acmart]  The \texttt{acmart}~\cite{acmart} style also redefines
  the \cmd{author} macro to capture individual authors, and by
  ordering of macros, allows the capture of affiliations and ORCID IDs
  for individual authors.
\item[amsart] The journal series of the American Mathematical Society~\cite{amsart} uses
  physical mail address rather than affiliations, but they support multiple
  addresses per author and the option of specifying a different ``current address''
  from where the research was done. They do not provide a field for
  ORCID IDs or ROR IDs. They also allow a second class of author called
  ``contributor'', where they are given credit for only a portion of the
  paper.
\item[elsarticle] The journal class by Elsevier also captures some metadata
  in a structured way. They use one \cmd{author} command per author, and they
  allow multiple footnote marks to be associated with each author. Their
  construction is nearly ideal, since each author can have multiple affiliations,
  and affiliations need not be repeated since they each have their own reference
  number. Moreover, affiliations take different attributes, so they are extensible
  to specify things like ROR IDs in the future.
\end{description}
Each of these provides some advance in capturing the metadata about an
article, but none of them rise to the level of expressiveness
contained in something like JATS. Moreover, we are unaware of any that
have attempted to provide functionality for a publishing workflow by extracting
the metadata from the \LaTeX.

\section{The \texttt{iacrcc} \LaTeX\ and \BibTeX\ styles}
Building on what we have learned from previous efforts, we have
designed a new document class called
\texttt{iacrcc}\footnote{May be downloaded from \aurl{publish.iacr.org/iacrcc}} 
that allows us to capture as much metadata as possible
from a document. This may be used with either \BibTeX\ with our own
\texttt{iacrcc.bst} style, or may be used with the \texttt{biblatex} package.
These files are designed to be used in a publishing
workflow to produce metadata in several different formats.  Not only
do they produce metadata to go back into PDF, but they also produce a
plain text version of metadata that can be easily processed for other
purposes like DOI registration.  We capture a broad range of metadata,
including alternate titles, author names, surnames, ORCID IDs,
affiliations with ROR IDs and addresses, and abstract. An example of author
metadata for \texttt{iacrcc} is given in Figure~\ref{fig:metadata}. Some of this
metadata can be expressed in the base schemas of XMP, but others
require extension schemas from JATS.

A few other steps are included in the implementation of
the \cmd{addauthor}, \cmd{title}, and \cmd{affiliation} macros to make
sure that they capture text metadata. With a few exceptions such as
accents, most macros are forbidden in the body of \cmd{title}.  Titles
are allowed to contain macros within inline mathematics, but if the
author wishes to use other macros in the title, then they must also
provide the \texttt{plaintext} argument that contains only text.  This
is similar to the requirement to provide a running title in the event
that a title is too long.  Footnotes in author names must be provided
with a \texttt{footnote} option to \cmd{addauthor}, and the use
of \cmd{thanks} is forbidden.

\begin{figure}[ht]
\begin{Verbatim}
\title[running={Emojex documentation},
       onclick={https://example.com/emo},
       subtitle={Emojis in LaTeX},
       plaintext={Emojex: use of emojis in LaTeX},
      ]{Emojex: use of emojis in \LaTeX}
\addauthor[orcid={0000-0002-0599-0192},
           inst={1,2},
           onclick={https://www.madmagazine.com/}
           email={fester@example.com},
           surname={Bestertester}
          ]{Fester Bestertester}
\addauthor[orcid={0000-0001-7890-5430},
           inst={2},
           footnote={Thanks mom!},
           surname={McCurley}
          ]{Kevin S. McCurley}
\affiliation[ror=044t1p926,
             city={New York},
             country={United States}]{MAD}
\affiliation[country={United States}]{Self}
\addfunding[crossref=100011047,
            grantid={A-1234},
            country={Canada}
           ]{AGE-WELL}
\end{Verbatim}
\caption{Sample metadata entry in \texttt{iacrcc.cls}\label{fig:metadata}}
\end{figure}

\subsection{How it works}
The workflow for an author consists of the usual multiple rounds of running
\texttt{latex}, \texttt{bibtex} or \texttt{biber}, followed by two
more runs of \texttt{latex}. The output from this is a PDF file
with XMP metadata, but also a file \cmd{jobname}\texttt{.meta} file
that contains all metadata in a structured format. The \texttt{.meta}
file is written with macros using \cmd{write} macros.

The structure of the \cmd{jobname}\texttt{.meta} is similar to
YAML. We thought about attempting to write YAML or JSON or XML format,
but each output format has its own set of special characters and
encoding requirements that are complicated to achieve in \LaTeX{}.
It was easier to write python code to parse our output format than
it is to write \LaTeX\ code to produce one of the more common formats.
This python code is included in the repository for the \texttt{iacrcc}.\footnote{See the github repository at \aurl{github.com/IACR/latex}.}

The basic metadata from the paper is written to the \texttt{.meta}
file using macros from the \texttt{iacrcc.cls} file. The citation
information is written into the \texttt{.meta} file in one of two
different ways, depending on whether the author chooses to use
\BibTeX\ or \texttt{biblatex}.  Both methods produce a \texttt{.bbl}
file that can be supplied in a journal workflow as a substitute for the
original \BibTeX\ files, (which may be huge and contain references
that are not used). Using either \BibTeX\ or \texttt{biblatex}, the
processing of the \texttt{.bbl} file results in the production of
citation records to appear in the \texttt{.meta} file through the use
of \cmd{write} macros. The \cmd{write} macros are implemented in the
\texttt{iacrcc.cls} file for \texttt{biblatex}, and are implemented in
the \texttt{iacrcc.bst} file for \BibTeX{}.  In both cases, the
\texttt{.bbl} file ends up containing a structured form of the
citations. This allows us to follow the standard practice of
publishers to only require authors to submit their \texttt{.bbl} file
rather than their entire \BibTeX\ file.

\subsection{The submission pipeline}
When the authors submit their paper, they need only submit their
\LaTeX\ source file(s), including the \BibTeX\ file they used.
The submission form is minimal, since all
metadata is included in the \LaTeX\ and \BibTeX\ files themselves. We
simply capture an authenticated \texttt{paperid} and require the
submitting author to supply an email address for the contact author.
We derive the DOI from the \texttt{paperid}, and inject it into the PDF
during compilation along with the acceptance date and received date.

Once the authors upload their \LaTeX\ sources, our server runs
\texttt{latexmk} within a docker container containing an instance of
texlive. We chose to use \texttt{latex} because we found it
impossible to make \texttt{pdflatex} produce UTF-8 output files from
\cmd{write}. The server validates that the sources were compiled, and
provides reports back to the author in case of any errors.

Once the document successfully compiles, then we run a python script
to process the \texttt{.meta} file, creating metadata in XMP, JATS,
JSON, and crossref formats. The JSON format is convenient for
immediately publishing the article on the web. The crossref format may be
used to register the paper with a DOI.

If the author is satisfied with the metadata and PDF produced from
compiling, then the paper moves to the next step of copy editing. In
our pipeline this step is fairly limited. We recognize that editors
may perform additional tasks such as grammar checking, punctuation
checking, or spell correction, but we believe that having to deal with
the metadata should not consume the time of a human.  While we have
attempted to address the issue of metadata handling, we believe that
copy editing remains as the biggest obstacle for lowering the cost of
open access publishing.

Once the paper is given final approval by the copy editor, the paper
may be published without need for a human to handle any of the metadata.
At the time the paper is published, the DOI is registered.

We have described the workflow using \texttt{latex}, but we can
also produce HTML output using {\TeX4ht} or {\LaTeX{ML}} or another
converter. We plan to investigate this once we have enough experience
with the PDF pipeline.

\section{Summary}
We believe that \LaTeX\ can be used to simplify the processing of
metadata in the publishing process, and we have developed a document
class that we hope will greatly improve our the quality of our
metadata. By using this approach, we believe that it should be
possible to streamline the publishing workflow of an open access
journal running on a low budget.

\section*{Acknowledgements}
The authors would like to thank Gaëtan Leurent and other contributors
to the \texttt{iacrtrans} document class that was used as the starting
point of this project. The authors would also like to thank Enrico
Gregorio and David Carlisle for answering questions about the inner
workings of \LaTeX.  We also wish to express our sincere thanks to
the \LaTeX\ team for their valuable recommendation on
using \cmd{protected@write}.

\bibliography{metadata}

\end{document}